\documentclass[secnumarabic,aps,prd,floatfix,nofootinbib,showkeys,twocolumn,showpacs,10pt,superscriptaddress]
{revtex4-1}

\usepackage{times}
\usepackage{amsfonts}
\usepackage{amssymb}%% The amssymb package provides various useful mathematical symbols
\usepackage{amsmath}
\usepackage{wasysym}
\usepackage{natbib}

\usepackage{graphicx}% Include figure files
\usepackage[tight,footnotesize]{subfigure}
\usepackage{enumerate}
\usepackage[inline]{enumitem}%\usepackage{enumitem}
\usepackage{moreenum}
\usepackage[utf8]{inputenc}
\usepackage[english,greek]{babel}

\usepackage{rotating}

\usepackage{xcolor}

\usepackage{setspace}%used to allow caption in table use the whole page size

\usepackage[bookmarks=true,colorlinks=true,citecolor=blue,linkcolor=blue,urlcolor=magenta]{hyperref}
\usepackage[normalem]{ulem}
\usepackage[most]{tcolorbox}
\usetikzlibrary{patterns}
\pgfdeclarepatternformonly{mystrikeout}{\pgfqpoint{-1pt}{-1pt}}{\pgfqpoint{11pt}{11pt}}{\pgfqpoint{10pt}{10pt}}%
{
  \pgfsetlinewidth{0.4pt}
  \pgfpathmoveto{\pgfqpoint{0pt}{0pt}}
  \pgfpathlineto{\pgfqpoint{10.1pt}{10.1pt}}
  \pgfusepath{stroke}
}
\newtcolorbox{tcbstrikeout}{breakable,
 enhanced jigsaw,
 opacityback=0,
 parbox=false,
 boxrule=0mm,
 top=0mm,bottom=0pt,left=0pt,right=0pt,
 boxsep=0pt,
 frame hidden,
 finish={\fill[pattern=mystrikeout] (frame.north west) rectangle (frame.south east);}
}

\def\aap{A\&A}
\def\aj{AJ}
\def\apj{ApJ}
\def\apjl{ApJL}
\def\apjs{ApJS}
\def\apss{Astrophysics and Space Science}
\def\jcap{JCAP}
\def\mnras{MNRAS}
\def\na{New Astronomy}
\def\nar{New Astronomy Reviews}
\def\pasp{PASP}
\def\physrep{Physics Reports}

\newcommand{\Mov}[1]{{\color{black}{#1}}}

\begin{document}

\title{Cluster density slopes from Dark Matter-Baryons Energy Transfer}

\author{Antonino Del Popolo}
\email{adelpopolo@oact.inaf.it}

\selectlanguage{english}%

\affiliation{Dipartimento di Fisica e Astronomia, University Of Catania, Viale Andrea Doria 6, 95125, Catania, Italy}

\affiliation{Institute of Astronomy, Russian Academy of Sciences, 
%,~\\
119017, Pyatnitskaya str., 48 , Moscow}

% \affiliation{INFN sezione di Catania,
% %~\\
% Via S. Sofia 64, I-95123 Catania, Italy}

\author{Morgan~\surname{Le~Delliou}}
\affiliation{Institute of Theoretical Physics, School of Physical Science and Technology, Lanzhou University, No.222, South Tianshui Road, Lanzhou, Gansu 730000, China}
\affiliation{Instituto de Astrof\'isica e Ci\^encias do Espa\c co, Universidade de Lisboa, Faculdade de Ci\^encias, Ed. C8, Campo Grande, 1769-016 Lisboa, Portugal}
%\affiliation{Universit\'e de Paris, APC-Astroparticule et Cosmologie (UMR-CNRS 7164), 
%Batiment Condorcet, 10 rue Alice Domon et L\'eonie Duquet, F-75205 Paris Cedex 13, France.
% F-75006 Paris, France}
\email[Corresponding author: ]{(delliou@lzu.edu.cn,) Morgan.LeDelliou.ift@gmail.com}
\affiliation{Lanzhou Center for Theoretical Physics, Key Laboratory of Theoretical Physics of Gansu Province, Lanzhou University, Lanzhou, Gansu 730000, China}
 \author{Maksym ~\surname{Deliyergiyev}}
 \affiliation{D\'epartement de Physique Nucl\'eaire et Corpusculaire, University of Geneva, CH-1211 Geneve 4, Switzerland}
 \email{maksym.deliyergiyev@unige.ch}

\label{firstpage}

\date{\today}

\begin{abstract}
In this paper, we extend  previous works on 
the relation between mass and the
inner slope in dark matter density profiles. We calculate that relation in the mass range going
from dwarf galaxies to cluster of galaxies. This was done thanks to a modeling of energy transfer via SN and AGN feedback, as well as via dynamical friction of baryon clumps. We show that, in the mass range above galaxy masses 
(Groups and
clusters), the inner slope-mass relation changes its trend. It flattens (towards less cuspy profile) around masses corresponding
to groups of galaxies and steepens again for large galaxy cluster masses. The flattening is
produced by the AGN outflows (AGN feedback). The one-$ \sigma$ scatter on $\alpha$ is approximately constant in all
the mass range ($\Delta\alpha\simeq 0.3$). 
This is the first paper extending the inner density profile slope-mass relationship to clusters of galaxies, accounting for the role of baryons. The result 
can be used to obtain a complete density profile, 
also 
taking baryons
into account. Such 
kind of density profile 
was previously
only 
available
for galaxies.
\end{abstract}

%\pacs{98.52.Wz, 98.65.Cw}
%\pacs{97.60.Jd, 97.10.Nf, 97.10.Pg, 95.35.+d, 26.60.-c}
%\keywords{Dwarf galaxies; galaxy clusters; missing satellite problem}

\maketitle

%%%%%%%%%%%%%%%%%%%%%%%%%%%%%%%
%%%%%%%%%%%%%%%%%%%%%%%%%%%%%%%
\section{Introduction}

The content of the Universe is clearly not reduced to ordinary baryonic matter, as seen in the gravitational effects at the cosmological level \citep{Planck2016}, mixed with astrophysical scale effects \citep{Bertone2005,DelPopolo2014}, where observations strongly point towards non-baryonic mass/energy domination of mass content by a clustering component coined dark matter (DM). Cosmological observations of the Universe's accelerated expansion \citep{Riess1998,Perlmutter1999} further indicate overall domination by the component responsible for this acceleration called dark energy (DE). 

The baryonic, DM and DE densities are measured to make up, respectively, $4.9\%$, $26.4\%$ and $68.7\%$ of our Universe.

Our most successful model for such universe, based on the big-bang cosmology, uses the cosmological constant $\Lambda$ as DE and describes the rest with five further parameters. Designated as the $\Lambda$CDM (Cold DM with $\Lambda$) model, this paradigm, although it was very successful to explain many phenomena, as found in numerous studies including \citep{DelPopolo2007,Komatsu2011,Hinshaw2013,DelPopolo2013,Planck2014,DelPopolo2014}, remains plagued with unexpected discrepancies compared with specific observations, as well as theoretical challenges: the deep theoretical challenges include what is known as the cosmological constant problems \citep{Weinberg1989,Astashenok2012}, related also to the unknown natures of DE \citep{DelPopolo2013a,DelPopolo2013b,DelPopolo2013c} and DM. From an observational perspective, anomalies accumulate at the large scales, in tensions of the Cosmic Microwave Background (CMB) Planck 2015 data with the Hubble parameter measured in type Ia SuperNovae (SNIa, or simply SN) \cite{Bolejko:2017fos}, with the CFHTLenS  weak lensing \citep{Raveri2016}, or with its $\sigma_8$ values \citep{Macaulay2013}. At small scales, they are embodied within the so-called "small scale problems" \citep[galactic, and centre of galaxy clusters scales, discussed, e.g., in][]{DelPopolo2000,DelPopolo2002a,DelPopolo2012c,Newman2013a,Newman2013b,DelPopolo2014,DelPopolo2017,DelPopolo:2018wrz}. They comprise 
\begin {enumerate*} [label=\itshape\alph*\upshape)]
\item the anomalous gap between observations and N-body simulations predictions on the number of galactic subhaloes  \citep[e.g.][]{Moore1999}; \item the so-called Too-Big-To-Fail (TBTF) problem: simulated haloes produce too many, too massive and dense subhaloes, that cannot be disrupted to explain their absence in observations \citep*{BoylanKolchin2011,BoylanKolchin2012}. \\

Those two problems found a proposal for a unified solution, based on the effect of baryons within the haloes' inner 
parts \citep{Zolotov2012,DelPopolo2014b}.\\ \item the Cusp/Core problem remains the most persistent of the $\Lambda$CDM paradigm problems \citep{Moore1994,Flores1994} and points at the inconsistency between LSBs and dwarf galaxies observed flat density 
profiles and the cuspy profiles produced in N-body simulations, e.g. 
the Navarro-Frenk-White (NFW) profile \citep[][]{Navarro1996,Navarro1997,Navarro2010}.
\end {enumerate*} 

This paper will focus on the so-called Cusp/Core problem. The density profiles from simulations already are subjects of discussions, focussing on the slopes for the inner region of haloes: the NFW profile predicts an 
inner profile characterised by density $\rho \propto r^\alpha$, with $\alpha=-1$, \cite{Moore1998} and \cite{Fukushige2001} produced even steeper profiles, with $\alpha=-1.5$, while other works encounter object and/or even mass dependent inner slopes
\citep{Jing2000,Ricotti2003,Ricotti2004,Ricotti2007,DelPopolo2010,Cardone2011b,DelPopolo2011,DelPopolo2013d,
DiCintio2014}. The Einasto profile, flattening towards the 
centre to $\alpha\simeq -0.8$ \citep{Stadel2009} seems to provide a better 
fit to simulations \citep{Gao2008}. The claim of universal density profiles have been contradicted in \cite{Polisensky2015}, where their initial linear density perturbation power spectra determine their shape, which also depend on their mass. This mass dependence agrees with previous works and with possible cores development in the warm DM (WDM) paradigm, however not 
significantly enough to explain observations.

In this debated context for the inner slope of haloes, the Cusp/Core problem resides in that dissipationless N-body simulations smallest predicted inner slopes exceed those obtained from SPH simulations 
\citep{Governato2010,Governato2012}, semi-analytical models 
\citep{DelPopolo2009,Cardone2012,DelPopolo2012a,DelPopolo2012b,DelPopolo2014a}, or observations \citep{Burkert1995,deBlok2003,Swaters2003,KuziodeNaray2011,Oh2011a,Oh2011b}. 

Although the Cusp/Core discussion started from galaxy scale haloes, it also has impact at galaxy clusters scales. Even though clusters total mass profiles agree with NFW predictions 
\citep{Sand2002,Sand2004,Newman2013a,Newman2013b,DelPopolo:2019syg}, lensing and kinematics constraints applied in relaxed clusters' central cD galaxies (Brightest Central Galaxies, BCG) found flatter DM profiles than the NFW.

The simple dynamical structure (bulgeless disks) of dwarf galaxies and their DM domination with low baryon fraction \citep{deBlok1997} made them widely used in the Cusp/Core debate, as the determination of the inner density structure of larger, high surface brightness (HSB) objects is more complicated and the universality of galaxies' cored nature is not definitely established: some authors claim HSBs are cored \cite{Spano2008} when others differ \citep[e.g.,][]{Simon2005,deBlok2008,DelPopolo2012c,DelPopolo2013d,Martinsson2013}. While low luminosity galaxies, 
$M_B>-19$,  in the THINGS sample, tend to follow isothermal (ISO) profiles, cuspy or cored profiles describe equally well its galaxies with $M_B<-19$. The inner profile of dwarfs galaxies also varies \cite{Simon2005}: among NGC 2976, 4605, 5949, 5693, and 6689, it ranges from 0 (NGC2976) to -1.28 (NGC5963). The confusion increases when noting that similar techniques on the same object yield different results: for, e.g., NGC2976, \cite{Simon2003} obtained $-0.17<\alpha<-0.01$ for the DM slope, while 
\cite{Adams2012} got $\alpha=-0.90 \pm 0.15$, \cite{Adams2014} traced $\alpha=-0.53 \pm 0.14$ with stars, or $\alpha=-0.30 \pm 0.18$ was derived using gas by \citep{Adams2014}.

This discussion reveals the difficulties associated with galaxies inner slope determination, including for dwarfs. It shows the existence of a range of slopes and the lack of agreement on their distribution despite recent kinematic maps \citep{Simon2005,Oh2011b,Adams2014}.

The confusion increases further for smaller masses (e.g. dwarf 
spheroidals (dSphs)) and larger masses (e.g., spiral galaxies), where stars dominate\footnote{See Section~\ref{sec:Results} for a wider discussion.}, when biases enter models \citep*{Battaglia2013} and yields opposite outcomes.

Evaluation of central slopes of dSphs can employ different methods: as mass and stellar orbits anisotropy are degenerate in the spherical Jean's equation model, its results strongly depend on the model's assumptions \citep{Evans2009}; a similar drawback also plaguing the maximum likelihood method applied to Jean's model parameter space \citep{Wolf2012,Hayashi2012,Richardson2013}. Schwarzschild modelling found cored profiles for the Fornax and Sculptor profiles \citep{Jardel2012,Breddels2013,Jardel2013b,Jardel2013a}, in agreement with multiple stellar populations methods that can measure central slopes at $\simeq$ 1 kpc (Fornax) and even $\simeq$ 500 pc (Sculptor) \citep{Battaglia2008,Walker2011,Agnello2012,Amorisco2012}. However the Schwarzschild model applied to Draco found a cusp \citep{Jardel2013a}. 
In general, there is not consensus on the inner structure of dSphs. A recent paper \cite{Hayashi:2020jze} found different, and cuspier, halo density profiles than previous estimates. Similarly, \cite{Shao:2020tsl} showed that for Fornax, considered for a long time to harbour a 1 kpc core, a
cuspy dark matter halo is probably not excluded.
%Generally speaking there is not consensus on the inner
%structure of dSphs. A recent paper \cite{Hayashi:2020jze} found different halo
%density profile, and more cuspy than previous estimates.
%Similarly \cite{Shao:2020tsl} showed that probably that Fornax considered for a long time characterized by a core of 1 kpc, a
%cuspy dark matter halo is not excluded.

Although dSphs can thus either be cored or cusped, their DM dominated dynamics should yield cuspy central profiles, at least for smaller masses.

Given observations, two kinds of approaches could solve the Cusp/Core problem:
\begin {enumerate} [label=\arabic*\upshape)]
 \item cosmological solutions, comprising\begin{enumerate}
                                          \item modified small scale initial spectrum \citep[e.g.][]{Zentner2003}, 
                                          \item various DM particles nature 
\citep{Colin2000,Goodman2000,Hu2000,Kaplinghat2000,Peebles2000,SommerLarsen2001}, 
\item modified gravity theories, e.g., $f(R)$ \citep{Buchdahl1970,Starobinsky1980}, $f(T)$ 
\citep[see][]{Bengochea2009,Linder2010,Dent2011,Zheng2011} or MOND \citep{Milgrom1983b,Milgrom1983a}.
                                         \end{enumerate}
 \item \label{enu:AstroSol} Astrophysical solutions, that reduce galaxies' inner density from DM component expansion induced by some "heating" mechanism, such as\begin{enumerate}
                                                                                                                                             \item ``supernovae feedback`` cusp flattening (SNF)
\citep{Navarro1996a,Gelato1999,Read2005,Mashchenko2006,Mashchenko2008,Governato2010,Governato2012,Onorbe:2015ija,El-Badry2017,Fitts:2016usl},
\item "dynamical friction from baryonic clumps" (DFBC) \label{enu:DFBC}
\citep{ElZant2001,ElZant2004,Ma2004,Nipoti2004,RomanoDiaz2008,RomanoDiaz2009,DelPopolo2009,Cole2011,Inoue2011,
Nipoti2015}.
                                                                                                                                            \end{enumerate}

\end {enumerate}

We will concentrate here in the astrophysical solutions \ref{enu:AstroSol}, and in particular in the DFBC \ref{enu:DFBC}, discussed further in 
the following sections.

%Concerning the SNF, it is important to remember that cos-
%mological simulations with lower density thresholds for star
%formation, e.g. APOSTLE and Auriga \citep{Bose:2018oaj}, does not pro-
%duce dark matter cores. Then the SNF is succesfull in some
%cases and less in others. For example, the
The SNF model has been studied in a large number of papers, and has been shown to be most effective for galaxies smaller that the Milky Way \citep{Navarro1996a,Governato2010,Governato2012,Teyssier2013,Chan:2015tna,Tollet:2015gqa}.

Although successful in some cases, the SNF model %The SNF model, although succesfull in some cases, 
is less so in others. Its effects depend, among other things, on the nature of star formation. For instance, for the SNF, cosmological simulations with lower density thresholds for star
formation, e.g. APOSTLE and Auriga \citep{Bose:2018oaj}, do not produce DM cores. Furthermore, the THINGS galaxies
\citep{deBlok2008,Walter2008} density profiles \citep{Oh2008,Oh2011a,Oh2011b} agreed with the \cite{Governato2010}
SNF model, while simulations of both disk galaxies and
dwarfs from \cite{TrujilloGomez2015}, including the SNF model along with feedback from massive stars radiation pressure, found radiation pressure to dominate SNF effects in
core formation, so SNF alone cannot form cores.
%that the role
%of radiation pressure is more important than that of SNF in
%core formation. Namely, SNF cannot form cores.
In general,
the SNF as possible solution to $\Lambda$CDM small scale problems
has been questioned in many works \citep{Ferrero2012,Penarrubia2012,GarrisonKimmel2013,GarrisonKimmel2014,Papastergis2015}.
Core formation is also influenced by the ratio of DM halo growth time to star formation time%timing of dark matter halo growth relative to star formation
. Mergers happening after core formation can rejuvenate a cusp %, cusp can reborn 
\citep{Onorbe2015}.

%Other papers claim an agreement between galaxies characteristics and simulations, as \citep{Katz:2016hyb}. This last paper tested the
% prediction of \cite{DiCintio2014}, which showed that when the ratio between
% the stellar mass $M_*$ , and that of the halo $M_{halo}$ is of the order or smaller than 0.01 %, energy from SNF is not enough
% to give rise to a core, while at larger values, SNF gives rise to
% an expansion of the dark matter giving rise to a core.
% The flattest profile forms when $M_* /M_{halo}\simeq 5 \times 10^{-3} $. For
% larger ratios, the stellar component formed in the center of
% the structure deepens the gravitational potential opposing the
% expansion generated by the SNF, originating cuspier profiles.
% \citep{Chan:2015tna} using the FIRE-1 suite, and \cite{Tollet:2015gqa} by means of NIHAO
% suite confirmed the \cite{DiCintio2014} result.
Other papers, such as \citep{Katz:2016hyb}, claim an agreement between galaxies characteristics and simulations. \citep{Katz:2016hyb} tested the
prediction of \cite{DiCintio2014} that claims core formation when the $M_{\ast}/M_{\rm halo}$ ratio between the stellar and halo masses is of the order or smaller than 0.01, while at larger values, SNF leads DM to expand, again tending to give rise to a core.
The flattest profile forms when $M_{\ast}/M_{\rm halo}\simeq 5 \times 10^{-3} $. For
larger ratios, the structure's central stellar component deepens the gravitational potential, opposing the
SNF driven expansion, resulting in a cuspier profiles.
\citep{Chan:2015tna}, using the FIRE-1 suite, and \cite{Tollet:2015gqa}, by means of the NIHAO
suite, confirmed the \cite{DiCintio2014} result.

% The work of \cite{Tollet:2015gqa} was extended by \cite{Maccio:2020svl} including the black
% hole feedback, and so determining the relation mass-slope for
% eight order in magnitudes in stellar mass.
% \cite{DelPopolo2016a} compared the capability of the SNF and he DFBC
% mechanisms to solve the Cusp/Core problem by comparing
% their theoretical predictions to observations of the inner slopes
% of galaxies with masses ranging from dSphs to normal spirals.
% It was shown that both mechanisms give similar results.
The work of \cite{Tollet:2015gqa} was extended by \cite{Maccio:2020svl} to include black
hole (BH) feedback, determining the mass-slope relation over eight order in magnitudes in stellar mass.
\cite{DelPopolo2016a} compared the ability of the SNF and he DFBC
mechanisms to solve the Cusp/Core problem through their theoretical predictions vs observations of the inner slopes of galaxies confrontation, with masses ranging from dSphs to normal spirals.
It found both mechanisms to give similar results.
% The DFBC has several achievements, also described in next
% section. To start with, the DFBC predicted the correct shape
% of galaxy
The DFBC achievements, summarised hereafter, are also described in Sec.~\ref{sec:Implementation}. It predicted the correct shape
of galaxy  
 density profiles \citep{DelPopolo2009,DelPopolo2009a} in agreement with \cite{Governato2010,Governato2012}
 SPH simulations. Similarly in the case of clusters, the density
 profile in \citep{DelPopolo2012a} matches predictions from the profiles simulated
 by \cite{Martizzi2012}. 
% As well several quantities, concerning clusters of
% galaxies, obtained in \citep{DelPopolo2012a} are in agreement with the observed
% quantities in \cite{Newman2013a,Newman2013b}.
% The results obtained by \cite{DiCintio2014}, showing a dependence of the
% inner slope on mass had already been found in \cite{DelPopolo2010}. A mass
% dependence of slope, as reported, had been found in the mass
% range going from dwarf galaxies to clusters \citep{DelPopolo2009,DelPopolo2010,DelPopolo2012a,DelPopolo2012b,DelPopolo2014}.
Furthermore, several galaxy clusters predictions from \citep{DelPopolo2012a} are in agreement with the observations in \cite{Newman2013a,Newman2013b}.
The work in \cite{DelPopolo2010} had already  found the dependence of the inner slope on mass claimed later by \cite{DiCintio2014}. Such slope-mass dependence had been reported over masses
ranging from dwarf galaxies to clusters \citep{DelPopolo2009,DelPopolo2010,DelPopolo2012a,DelPopolo2012b,DelPopolo2014}.
% The DFBC model found a series of other correlations between the inner slope and the ratio between baryonic content
% halo mass, and also angular momentum \citep{DelPopolo2012b},
% In agreement with \cite{Newman2013a,Newman2013b}, there are correlations between the
% inner slope and
The DFBC model found a series of additional correlations such as between \\
\begin{enumerate*}[label=\itshape\alph*\upshape)]
\item inner slope and
\begin{enumerate*}[label=\itshape\arabic*\upshape)]
 \item the baryon to halo mass ratio $M_{\rm b}/M_{500}$\footnote{Recall $R_{500}$ encloses 500 times the critical density and a mass $M_{500}$.}%; its correlation with the mass ratio doubles into one with the ratio at 
% $z=0$ and one at $z_{\rm initial}$, noted $M_{\rm b, in}/M_{500}$ 
% \citep[see figures 2, 4, in][]{DelPopolo2012b}. Here we refer to $M_{\rm b, in}$ as the protostructure initial gas mass, and to $M_{\rm b}$ as the final total baryonic mass, i.e. $M_{\rm b}= M_{\rm gas+stars}$. Note that the DFBC clusters inner slope correlation with $M_{\ast}/M_{\rm halo}$ mimics SPH simulations, although they use a different mechanism and model galactic haloes.\\
 \item angular momentum\footnote{Larger mass structure collapse is reduced since their acquired angular momentum follows 
the peak height in inverse proportion \citep{DelPopolo1996,DelPopolo2009}.}\\
\end{enumerate*}
as seen in \citep{DelPopolo2012b}, while, in agreement with \cite{Newman2013a,Newman2013b}, the inner slope also correlates with
\begin{enumerate*}[resume*]
\item  the Brightest Cluster Galaxy 
(BCG) mass, 
\item the core radius $r_{\rm core}$,
\item the effective radius $R_{\rm e}$,  and\\
\end{enumerate*}
\item between the DM dominated mass inside 100 kpc, and the mainly baryonic \citep{DelPopolo2014c} mass inside 5 kpc.
\end{enumerate*}

Ref.~\cite{Tollet:2015gqa} confirmed the results of \cite{DiCintio2014}, and extended it to redshift
z = 1. The FIRE-2 galaxy formation physics simulated 54 galaxy halos, which CDM density profiles were analyzed in \cite{Lazar:2020pjs}. Ref.~\cite{Maccio:2020svl} added 46 new high resolution simulations of massive galaxies, including BH feedback, to the work of \cite{Tollet:2015gqa}.
This allowed to trace the DM halo inner slope dependence from galaxies to groups of galaxies.
% \cite{Tollet:2015gqa} confirmed the results of \cite{DiCintio2014}, and extended it to redshift
% z = 1. In \cite{Lazar:2020pjs}, the cold dark matter density profiles of 54
% galaxy halos simulated with FIRE-2 galaxy formation physics
% were analyzed. \cite{Maccio:2020svl}, extended the previous work by \cite{Tollet:2015gqa} 
% paper by adding 46 new high resolution simulations of massive
% galaxies performed with the inclusion of Black Hole feedback.
% In this way they were able to trace the dependence of the inner
% slope of dark matter halo from galaxies to groups of galaxies.

In this context, \cite{DelPopolo2010} showed the halo density profiles inner slope of spiral galaxies depend on their mass. That result was extended in \cite{DelPopolo2016a} to spheroidal galaxies. For dSphs with baryonic mass smaller than $10^9 M_\odot$, the DM halo density profile was shown to steepen towards smaller masses. The slope shows % up to 
a maximum flattening at $\simeq10^9 M_\odot$, before steepening again for larger masses. A similar dependence was reported by \cite{DiCintio2014}, which work was extended by \cite{Maccio:2020svl,Tollet:2015gqa,Lazar:2020pjs}. The NIHAO and FIRE results were compared in \cite[Fig. 6]{Hayashi:2020jze} for the slope-mass relation \cite[also see][Fig. 2]{Lazar:2020pjs}. The \cite{Lazar:2020pjs} slope-mass relation exclusively concerned galaxies, while only \cite{Maccio:2020svl} extended it to groups of galaxies. In this paper, we aim to extend that study to galaxy clusters, including AGN feedback effects. That purpose will lead us to improve and extend the \cite{DelPopolo2016a}
model, using the \cite{DelPopolo:2018wrz} model for AGN feedback. Building a dwarf galaxies-cluster size halo mass range, including DM and baryonic effects, will enable for the first time the construction of a mass-dependent DM density profile set. Recall that the \cite{DiCintio2014} profiles only concerned galaxies.

This paper aims to extend %The aim of this paper is that of extending 
some of the results of \citep{DelPopolo2016a}, namely those related to the slope-mass relationship. %In 
Ref.~\citep{DelPopolo2016a}, among other results, %we 
showed how the slope of the inner DM %dark matter 
density profile depends on baryonic and halo mass. 

% \Mov{{\bf[I would replace the following with my version]}The slope shows a maximum flattening at stellar masses $\simeq 10^8 M_\odot$. At smaller masses the steepening reaches a value of $\alpha \simeq -0.6$ at $M_{\rm star} \simeq 10^4 M_\odot$. This shows that the steepening at small masses is smaller that in the case of models based on supernovae feedback like \citep{DiCintio2014,Tollet:2015gqa}. In other words, the model predicts that dwarf galaxies are less cuspy than what predicted by \citep{DiCintio2014,Tollet:2015gqa}. This result is very important because comparing with the observed slopes of dwarf, and ultra faint galaxies one can understand what is the mechanism giving rise to the inner structure of those galaxies. Dwarf, or ultra-faint galaxies with a almost cored profile means that the DFBC mechanism is the responsible of the core formation, while a cuspy profile implies that supernovae feedback has the main role in the cusp formation.}
The mass range studied went %was that 
from dwarf galaxies to galaxies similar to our Galaxy. The present paper extends %In the present paper we want to extend 
that range to %the 
clusters of galaxy masses. To date, halo density profile taking baryons into account are only available for %we have just the halo density profile taking into account of baryons, to the 
galaxies with mass similar to the Milky Way's.  

The importance of the present results lies in the possibility it opens to compare the observed slopes of dwarf, and ultra faint galaxies, and thus discriminate which mechanism gives rise to the inner structure of those galaxies. Indeed, the slope of our model shows a maximum flattening at stellar masses $\simeq 10^8 M_\odot$. At smaller masses the steepening reaches a value of $\alpha \simeq -0.6$ at $M_{\rm star} \simeq 10^4 M_\odot$. Such steepening at small masses is smaller that in the case of models based on supernovae feedback, such as \citep{DiCintio2014,Tollet:2015gqa}. In other words, the model predicts that dwarf galaxies are less cuspy than predicted by \citep{DiCintio2014,Tollet:2015gqa}. Consequently, dwarf, or ultra-faint galaxies with a almost cored profile means that the DFBC mechanism is the responsible of the core formation, while a cuspy profile implies that supernovae feedback has the main role in the cusp formation.

The paper is organized as follows. In Sec.~\ref{sec:Implementation} we describe the model that we 
will use\Mov{, which implementation is summarised in Sec.~\ref{sec:Simulation}}. Sec.~\ref{sec:DFBC and SNF} discusses SNF and the DFBC mechanisms.
Secs.~\ref{sec:Results} and~\ref{sec:Conclusions} are devoted to results and conclusions, 
respectively. %Finally in the appendix \ref{sect:app} we discuss in detail the DFBC mechanism.
% \Mov{\bf [more detail??]}
% \Mcom{*******************************}
% \cite{DiCintio:2014xia}
%%%%%%%%%%%%%%%%%%%%%%%%%%%%%%%
%%%%%%%%%%%%%%%%%%%%%%%%%%%%%%%
\section{Theoretical Model}
\label{sec:Implementation}

This section recalls the model employed in this work. %After being proposed in \cite{DelPopolo2009,DelPopolo2009a}, this very significant improvement on the spherical collapse models \citep{Gunn1972,Bertschinger1985,Hoffman1985,Ryden1987,Ascasibar2004,Williams2004} includes the effects of  random angular momentum induced by random motion during the collapse phase of haloes \citep[e.g.,][]{Ryden1987,Williams2004}, of ordered angular momentum induced by tidal torques \citep[e.g.,][]{Ryden1988,DelPopolo1997,DelPopolo2000}, and was furthered to include the consequences of adiabatic contraction \citep[e.g.,][]{Blumenthal1986,Gnedin2004, Klypin2002,Gustafsson2006}, of dynamical friction between DM and baryonic gas and stellar clumps
The spherical collapse models \citep{Gunn1972,Bertschinger1985,Hoffman1985,Ryden1987,Ascasibar2004,Williams2004} was very significant improved in \cite[e.g.][]{DelPopolo2009,DelPopolo2009a} to include the effects of
\begin{enumerate}[leftmargin=*,label=$-$,noitemsep,partopsep=0pt,topsep=0pt,parsep=0pt]%\setlength\itemsep{-.1cm}[leftmargin=*,label=$\bullet$,noitemsep,partopsep=0pt,topsep=0pt,parsep=0pt]
 \item random angular momentum induced by random motion during the collapse phase of haloes \citep[e.g.,][]{Ryden1987,Williams2004},% of
 \item ordered angular momentum induced by tidal torques \citep[e.g.,][]{Ryden1988,DelPopolo1997,DelPopolo2000}, 
\end{enumerate}
and was furthered to include the consequences of
\begin{enumerate}[resume,leftmargin=*,label=$-$,noitemsep,partopsep=0pt,topsep=0pt,parsep=0pt]%\setlength\itemsep{-.1cm}
 \item adiabatic contraction \citep[e.g.,][]{Blumenthal1986,Gnedin2004, Klypin2002,Gustafsson2006}, %of
 \item dynamical friction between DM and baryonic gas and stellar clumps 
 % \citep{ElZant2001,ElZant2004,Ma2004,RomanoDiaz2008,RomanoDiaz2009,DelPopolo2009,Cole2011,Inoue2011, Nipoti2015}, of gas cooling, star formation, photoionization, supernova, and AGN feedback 
% \citep{DeLucia2008,Li2010,Martizzi2012} and of dark energy \citep{DelPopolo2013a,DelPopolo2013b,DelPopolo2013c}, and was further refined in  \cite{DelPopolo2014a,DelPopolo2016a,DelPopolo2016b}. This model produced results on the universality of density profiles \citep{DelPopolo2010,DelPopolo2011}, specific features of density profiles in galaxies \citep{DelPopolo2012a,DelPopolo2014} and clusters \citep{DelPopolo2012b,DelPopolo2014}, as well as a focus on galaxies inner surface-density \citep*{DelPopolo2013d}.
\citep{ElZant2001,ElZant2004,Ma2004,RomanoDiaz2008,RomanoDiaz2009,DelPopolo2009,Cole2011,Inoue2011, Nipoti2015}, %of
 \item gas cooling, star formation, photoionization, supernova, and AGN feedback \citep{DeLucia2008,Li2010,Martizzi2012} and %of 
 \item DE \citep{DelPopolo2013a,DelPopolo2013b,DelPopolo2013c}, 
\end{enumerate}
and was further refined in  \cite{DelPopolo2014a,DelPopolo2016a,DelPopolo2016b,DelPopolo:2016skd}. This model produced results on 
\begin{enumerate}[leftmargin=*,label=$\centerdot$,noitemsep,partopsep=0pt,topsep=0pt,parsep=0pt]
                                                                                                                    \item the universality of density profiles \citep{DelPopolo2010,DelPopolo2011}, 
                                                                                                                \item specific features of density profiles in 
                                            \begin{enumerate}[leftmargin=*,label=$\cdot$,noitemsep,partopsep=0pt,topsep=0pt,parsep=0pt]
                                                                                                                                                                                    \item galaxies \citep{DelPopolo2012a,DelPopolo2014} and
                                                                                                                                                                                    \item clusters \citep{DelPopolo2012b,DelPopolo2014},  
                                                                                                                                                            \end{enumerate}
                                                                                                                                                            \end{enumerate}
as well as a focus on                                                                                                                                                
\begin{enumerate}[resume,leftmargin=*,label=$\centerdot$,noitemsep,partopsep=0pt,topsep=0pt,parsep=0pt]
                                                                                                                                                        
\item galaxies inner surface-density \citep*{DelPopolo2013d}.
                                                                                                                                        \end{enumerate}
% the universality of density profiles \citep{DelPopolo2010,DelPopolo2011}, specific features of density profiles in galaxies \citep{DelPopolo2012a,DelPopolo2014} and clusters \citep{DelPopolo2012b,DelPopolo2014}, as well as a focus on galaxies inner surface-density \citep*{DelPopolo2013d}.

Although the model's key mechanism resides in dynamical friction (DFBC), we stress out that it includes all of the above effects (including SNF) that each only contribute at the level of some \%.

Its implementation occurs in several stages:
\begin{enumerate}
 \item The diffuse proto-structure of gas and DM expands, in the linear phase, to a maximum radius before DM re-collapses into a potential well, where baryons will fall.
\item In their radiative clumping, baryons form stars at the halo centre.
\item Then four effects happen in parallel
\begin{enumerate}
 \item the DM central cusp increases from baryons adiabatic contraction (at $z \simeq 5$ in the case of $10^9 M_{\odot}$ galaxies \citep{DelPopolo2009})
 \item the galactic centre also receive clumps that collapse from baryons-DM dynamical friction (DF)
 \item the DF energy and angular momentum (AM) transfer to DM  \citep[and stars][]{Read2005,Pontzen2012,Teyssier2013} results in an opposite effect to adiabatic contraction, and reduces the halo central density \citep{ElZant2001,ElZant2004}.
 \item the balance between adiabatic contraction and DF can result in heating cusps and forming cores, i.e. in dwarf spheroidals and spirals, while the deeper potential wells of giant galaxies keeps their profile steeper.
\end{enumerate}
\item The effect of DF %deepens from 
adds to that of tidal torques (ordered AM), and random AM. 
  \item \label{SNmech} Finally, the core further slightly (few percent) enlarges from the decrease of stellar density due to successive gas expulsion from supernovae explosions, and from the disruption of the smallest gas clumps, once they have partially converted to stars \citep[see][]{Nipoti2015}.
\end{enumerate}

\subsection{Model treatment of density profile}

Starting from a Hubble expansion, the spherical model of density perturbations expands linearly until reaching a turn-around maximum and reverting into collapse \citep{Gunn1977,Fillmore1984}. A Lagrange particle approach yields the final density profile
\begin{equation}\label{eq:dturnnn}
 \rho(x)=\frac{\rho_{\rm ta}(x_{\rm m})}{f(x_{\rm i})^3}
 \left[1+\frac{d\ln{f(x_{\rm i})}}{d\ln{g(x_{\rm i})}}\right]^{-1}\;,
\end{equation}
with initial and turn-around radius, resp. $x_{\rm i}$ and $x_{\rm m}(x_{\rm i})$, collapse factor $f(x_{\rm i})=x/x_{\rm m}(x_{\rm i})$, and turnaround density $\rho_{\rm ta}(x_{\rm m})$. The turn-around radius is obtained with
\begin{equation}
 x_{\rm m}=g(x_{\rm i})=x_{\rm i}\frac{1+\overline{\delta}_{\rm i}}
 {\overline{\delta}_{\rm i}-(\Omega_{\rm i}^{-1}-1)}\;, 
\end{equation}
where we used $\Omega_{\rm i}$ for the density parameter, and $\overline{\delta}_{\rm i}$ for the average 
overdensity inside a DM and baryons shell.

The model starts with all baryons in gas form with $f_{\rm b}=0.17\pm 0.01$ for the ''universal baryon fraction`` \citep{Komatsu2009} \citep[set to 0.167 in][]{Komatsu2011}, before star formation proceeds as described below.

Tidal torque theory (TTT) allows to compute the ''specific ordered angular momentum``, $h$, exerted on smaller scales from larger scales tidal torques  \citep{Hoyle1953,Peebles1969,White1984,Ryden1988,Eisenstein1995}, while the ''random angular momentum``, $j$, %proceeds from interaction with 
is related to %{\bf[why not proceeds from interaction with?Still a valid question]} 
orbits eccentricity $e=\left(\frac{r_{\rm min}}{r_{\rm max}}\right)$ \citep{AvilaReese1998}, obtained from the apocentric radius $r_{\rm max}$, the pericentric radius $r_{\rm min}$ and corrected from the system's dynamical state effects advocated by \cite{Ascasibar2004}, using the spherically averaged turnaround radius $r_{\rm ta}=x_{\rm m}(x_{\rm i})$ and the maximum radius of the halo $r_{\rm max}<0.1 r_{\rm ta}$
\begin{equation}
e(r_{\rm max})\simeq 0.8\left(\frac{r_{\rm max}}{r_{\rm ta}}\right)^{0.1}\;.
\end{equation}

These corrections to the density profile are compounded also with its steepening from the adiabatic compression following \cite{Gnedin2004} and the effect of DF introduced in the equation of motion by a DF force \citep[see][Eq. A14]{DelPopolo2009}.

\subsection{Effects of baryons, discs, and clumps}

The baryon gas halo settles into a stable, rotationally supported, disk, in the case of spiral galaxies. Their size and mass result from solving the equation of motion, and lead to a solution of the angular momentum catastrophe (AMC) \citep[Section 3.2, Fig. 3, and 4 of][]{DelPopolo2014}, obtaining realistic disc size and mass.

Notwithstanding stabilization from the shear force, Jean's criterion shows the appearance of instability for denser discs. The condition for this appearance and subsequent clump formation was found by Toomre \cite{Toomre1964}, involving the 1-D velocity dispersion $\sigma$,\footnote{$\simeq 20-80$ km/s in most clump hosting galaxies} angular velocity $\Omega$, surface density $\Sigma$, related to the adiabatic sound speed $c_s$, and the epicyclic frequency $\kappa$
\begin{equation}
Q \simeq \sigma \Omega/(\pi G \Sigma)=\frac{c_s \kappa}{\pi G \Sigma}<1\;.
\end{equation}
The solution to the perturbation dispersion relation $d \omega^2/d k=0$ for $Q<1$ yields the fastest growing mode  $k_{\rm inst}=\frac{\pi G \Sigma}{c_s^2}$ (see \cite{BinneyTremaine1987} or \cite[Eq. 6]{Nipoti2015}). That condition allows to compute the clumps radii in galaxies \citep{Krumholz2010}
\begin{equation}
 R \simeq 7 G \Sigma/\Omega^2 \simeq 1 {\rm kpc}\;.
\end{equation}
Marginally unstable discs ($Q \simeq 1$) with maximal velocity dispersion have a total mass three times larger than that of the cold disc, and form clumps 
$\simeq 10$ \% of their disk mass $M_d$ \citep{Dekel2009}.

%{\bf
Objects of masses few times $10^{10}~M_{\odot}$, found in $5 \times 10^{11} M_{\odot}$ haloes at $z \simeq 2$, are in a marginally unstable phase for $\simeq 1$~Gyr. Generally the main properties of clumps are similar to those found by \cite{Ceverino2012}.

In agreement with 
\cite{Ma2004,Nipoti2004,RomanoDiaz2008,RomanoDiaz2009,DelPopolo2009,Cole2011,Inoue2011,DelPopolo2014d,Nipoti2015}, 
energy and AM transfer from clumps to DM flatten the profile more efficiently in smaller haloes.
%}

% Clumps formation for marginally unstable discs ($Q \simeq 1$), with triple the cold disc mass and maximal velocity dispersion, can reach $\simeq 10$ \% of their disk mass $M_d$ \citep{Dekel2009}. At $z \simeq 2$, haloes with $5 \times 10^{11} M_{\odot}$ were found with $\simeq 10^{9} M_{\odot}$ clumps that formed for $\simeq 1$ Gyr in a few $\times 10^{10} M_{\odot}$ marginally unstable discs, while clump masses in dwarf galaxies remain $\simeq 10^5 M_{\odot}$. 
% 
% In general the model's clump characteristics (density, rotation velocity) agree with \cite[see their fig. 15, 16]{Ceverino2012}, and its smaller, earlier, haloes display more efficient clumps-DM energy and AM transfer, as found in \cite{Ma2004,Nipoti2004,RomanoDiaz2008,RomanoDiaz2009,DelPopolo2009,Cole2011,Inoue2011,DelPopolo2014d,Nipoti2015}

\subsubsection{Computing the clumps life-time}

Evidence for existence of the clumps produced by the model can be traced both in simulations \citep[e.g.,][]
{Ceverino2010,Perez2013,Perret2013,Ceverino2013,Ceverino2014,Bournaud2014,Behrendt2016}, and observations. High redshift galaxies have been found to contain clump clusters or clumpy structures that leads to call them chain galaxies \citep[e.g.,][]{Elmegreen2004,Elmegreen2009,Genzel2011}. The HST Ultra Deep Field encompasses galaxies with massive star-forming clumps \citep{Guo2012,Wuyts2013}, many at $z=1-3$ \citep{Guo2015}, some in deeper fields $z \apprle 6$ \cite{Elmegreen2007}.

Such clumpy structures are expected to originate from self-gravity instability in very gas-rich disc, from radiative cooling in the accreting dense gas
%where radiative cooling occurs while the dense gas accretes 
\citep[e.g.,][]{Noguchi1998,Noguchi1999,Aumer2010,Ceverino2010,Ceverino2012}. Their effect on halo central density depend crucially on the clump lifetime: should their disruption through stellar feedback still allow them sufficient time
%they have sufficient time, despite their disruption through stellar feedback, 
to sink to the galaxy centre, they can turn a cusp into a core. A clump's ability to form a bound stellar system is assessed through its stellar feedback mass fraction loss,
%loss mass fraction, 
$e_f$, and its formed stars
%star formation 
mass fraction, $\varepsilon=1-e_f$. Simulations and analytical models agree that most of the mass of
such %formed 
group of stars will remain bound for $\epsilon \geq 0.5$ \cite{Baumgardt2007}. The radiation feedback efficiency can be estimated, using 
\begin{enumerate*}[label=\itshape\alph*\upshape)]
 \item the dimensionless star-formation rate efficiency $\epsilon_{eff}=\frac{\dot{M_*}}{M/t_{ff}}$. This %, which 
is simply the ratio between %of 
free-fall time, $t_{ff}$, and the %to 
depletion time for a stellar mass $M_\star$. In %, in 
its reduced version it reads $\epsilon_{eff},_{-2}=\epsilon_{eff}/0.01$,\\% ,
\item the reduced dimensionless surface density $\Sigma_1=\frac{\Sigma}{0.1 g/cm^2}$, and
\item the dimensionless reduced mass $M_9=M/10^9 M_{\odot}$,\\
\end{enumerate*}

\vspace{-.4cm}\noindent to obtain the expulsion fraction $e_f=1-\varepsilon=0.086 (\Sigma_1 M_9)^{-1/4} \epsilon_{eff},_{-2}$ \citep{Krumholz2010}. Ref.~\citep{Krumholz2007} estimated, for a large sample of environments, densities, size and scales, that $\epsilon_{eff}\simeq 0.01$. Furthermore, $e_f=0.15$ and $\varepsilon=0.85$ for typical clumps with masses  $M\simeq10^9 M_{\odot}$. Therefore, the clump mass loss before they reach the centre of the galactic halo should be small. However, such conclusion and the expulsion fraction method are valid for smaller, more compact clumps in smaller galaxies. Such context only produces clumps that survive all the way to the centre.
%These results indicated that the clump mass loss before they reach the centre of the galactic halo should be small ($e=0.15$ and $\varepsilon=0.85$) for a large sample of environments, $\epsilon_{eff}= 0.01$ \citep{Krumholz2007}, densities, size scales, and for typical clump masses  $M=10^9 M_{\odot}$.  

Alternately, comparing a clump lifetime to its migration time to the centre, one can also obtain clump disruption. Migration time is the result of DF and TTT: for a $10^9 M_{\odot}$ clump, it yields $\simeq 200$ Myrs \citep[see Eq. 1 of][Eq. 18]{Genzel2011,Nipoti2015}. Coincidents %A similar 
expansion %timescale 
and migration timescales were % was 
computed from the  Sedov-Taylor solution \citep[Eqs. 8,9]{Genzel2011}.

Clump lifetime has been much studied. Ceverino \emph{et al.}, finding clumps in Jean's equilibrium and rotational support, from hydrodynamical simulations \citep{Ceverino2010}, construed their
%Hydrodynamical simulations \citep{Ceverino2010} construed, from their Jean's equilibrium and rotational support, a 
long lifetime ($\simeq 2 \times 10^8$ Myr). This agrees with several approaches: in local systems forming stars and coinciding with the Kennicutt-Schmidt law,  \citep{Krumholz2010} found such lifetimes. This is because as clumps retained gas, and formed bound star groups, they %and 
had time to migrate to the galactic centre. Simulations from \cite{Elmegreen2008} confirmed it. Other simulations with proper account of stellar feedback, e.g. non-thermal and radiative feedback mechanisms, %exhaustive account, e.g. non-thermal and radiative feedback mechanisms, of stellar feedback 
also obtained long-lived clumps reaching galactic centre \citep[SNF, radiation pressure, etc][]{Perret2013,Bournaud2014,Ceverino2013}. Finally, %while 
the same was obtained with any reasonable amount of feedback \cite{Perez2013}. The expansion, gas expulsion, and metal enrichment, time scales (respectively $>100$ Myrs, 170-1600 Myrs, and $\simeq 200$ Myrs) obtained by \cite{Genzel2011} to estimate clump ages also bring strong evidence for long-lived clumps. Lastly, comparison between similar low and high redshift clumps observations \citep[in radius, mass,][]{Elmegreen2013,Garland2015,Mandelker2015} supports clump stability.

\subsection{Model treatment of feedback and star formation}

Star formation, reionisation, gas cooling, and SNF in the model are built along \citep[Secs.~2.2.2 and~2.2.3]{DeLucia2008,Li2010}.

\begin{description}
 \item[Reionisation] acts for $z=11.5-15$ by decreasing the baryon fraction as
\begin{equation}
f_{\rm b, halo}(z,M_{\rm vir})=\frac{f_{\rm b}}{[1+0.26 M_{\rm F}(z)/M_{\rm vir}]^3}\;,
\end{equation}
\citep{Li2010}, using the virial mass,$M_{\rm vir}$, and the ``filtering mass'' \citep[see][]{Kravtsov2004}, $M_{\rm F}$. 
\item[Gas cooling] follows from a cooling flow model \citep[e.g.,][see Sect. 2.2.2]{White1991,Li2010}.
\item[Star formation] arises from gas conversion into stars when it has settled in a disk. The gas mass conversion into stars during a given time interval $\Delta t$, which we take as the disc dynamical time $t_{\rm dyn}$, is given by
\begin{equation}
 \Delta M_{\ast}=\psi\Delta t\;,
\end{equation}
where the star formation rate $\psi$ comes from the gas mass above the density threshold $n>9.3/{\rm cm^3}$ \citep[fixed as in][]{DiCintio2014} according to \citep[see][for more details]{DeLucia2008}
\begin{equation}
\psi=0.03 M_{\rm sf}/t_{\rm dyn}\;.
\end{equation}
\item[SNF] follows \cite{Croton2006}, where SN explosions inject energy in the system. This energy can be %, 
calculated from a Chabrier IMF \cite{Chabrier2003}, using \begin{itemize}[leftmargin=*,label=$-$,noitemsep,partopsep=0pt,topsep=0pt,parsep=0pt]
                                                           \item the disc gas reheating energy efficiency $\epsilon_{\rm halo}$,
                                                           \item the available star mass $\Delta M_{\ast}$, 
                                                           \item that mass conversion into SN measured with the SN number per solar mass as $\eta_{\rm SN}=8\times 10^{-3}/M_{\odot}$, and
                                                           \item the typical energy an SN explosion releases $E_{\rm SN}=10^{51}$ erg,
                                                          \end{itemize}
 to obtain
\begin{equation}
 \Delta E_{\rm SN}=0.5\epsilon_{\rm halo}\Delta M_{\ast} \eta_{\rm SN}E_{\rm SN}\;.
\end{equation}
This released energy from SNs into the hot halo gas in the form of reheated disk gas then compares with the reheating energy $\Delta E_{\rm hot}$ which that same amount of gas should acquire if its injection in the halo should keep its specific energy constant, that is if the new gas would remain at equilibrium with the halo hot gas. That %the 
amount of disk gas the SN and stars radiation have reheated, $\Delta M_{\rm reheat}$, %would acquire if it remained at equilibrium with the hot gas halo, that is keeping the halo gas specific energy constant. The radiations from stars, measured by their mass, produces the reheated gas mass and the latter is thus 
 since it is produced from stars radiations, is 
proportional to their mass
\begin{equation}
 \Delta M_{\rm reheat} = 3.5 \Delta M_{\ast}\;.
\end{equation}
Since
%To keep constant 
the halo hot gas specific energy %, which 
corresponds to the Virial equilibrium specific kinetic energy $\frac{V^2_{\rm vir}}{2}$, keeping this energy constant under addition of that %this 
reheated gas leads to define the equilibrium reheating energy as
\begin{equation}
\Delta E_{\rm hot}= 0.5\Delta M_{\rm reheat} %\eta_{\rm SN}E_{\rm SN}
V^2_{\rm vir}\;.
\end{equation}
The comparison with the actual energy of the gas injected from the disk into the halo by SNs %SN produced energy 
gives the threshold ($\Delta E_{\rm SN}>\Delta E_{\rm hot}$) beyond which gas is expelled, %and 
the available energy to expel the reheated gas, and thus the amount of gas ejected from that extra energy
\begin{equation}
 \Delta M_{\rm eject}=\frac{\Delta E_{\rm SN}-\Delta E_{\rm hot}}{0.5 V^2_{\rm vir}}\;.
\end{equation}\\
Contrary to SNF based models such as \cite{DiCintio2014}, our mechanism for cusp flattening initiates before the star formation epoch. Since it uses %, with 
a gravitational energy source, it %and 
is thus less limited in available time and energy. Only after DF shapes the core can Stellar and SN feedback occurs, which then disrupt gas clouds in the core \citep[similarly to][]{Nipoti2015}.
%%%%%%%%%%%%%%%%%%%%%%%%%%%%%%%%%%%%%%%%%%
\item[AGN feedback] occurs when a central Super-Massive-Black-Hole (SMBH) is formed. We follow the prescriptions of \cite{Martizzi2012,Martizzi2012a}, modifying the \cite{Booth2009} model for SMBH mass accretion and AGN feedback: a seed $10^5~M_{\odot}$ SMBH forms when stellar density, reduced gas density ($\rho_{gas}/10$) and 3D velocity dispersion exceed the thresholds $2.4 \times 10^6 M_{\odot}/{\rm kpc}^3$ and 100 $km/s$, which then accretes. Significant AGN quenching starts above $M\simeq 6 \times 10^{11} M_{\odot}$ \citep{Cattaneo2006}.
\end{description}

\subsection{Model robustness}

We point out that the model demonstrated its robustness in various behaviours:

\begin{enumerate}[label=\greek*.]
        \item the cusp flattening from DM heating by collapsing baryonic clumps predicted for galaxies and clusters is in agreement with following studies %it predicted, from collapsing baryonic clumps DM heating, galaxies cusp flattening since 2009, and clusters' since 2012, in agreement with studies of
        \citep{ElZant2001,ElZant2004,RomanoDiaz2008,RomanoDiaz2009,Cole2011,Inoue2011,Nipoti2015}. A comparison with \cite{Governato2010}'s  SPH simulations was made in \citep[][Fig. 4]{DelPopolo2011}.
        \item it aforetime predicted the correct shape of galaxies density profiles  \citep{DelPopolo2009,DelPopolo2009a}, ahead of SPH simulations of \cite{Governato2010,Governato2012}, and of clusters density profiles \citep{DelPopolo2012b}  anteriorly of \cite{Martizzi2013}. \footnote{Note that \cite{Governato2010,Governato2012} and \cite{Martizzi2013} adopted different dominant mechanisms.}
        \item it aforetime predicted the halo mass dependence of cusps inner slope \cite[Fig. 2a solid line]{DelPopolo2010} beforehand the similar result in the non-extrapolated part of the plot in \cite[Fig. 6]{DiCintio2014}, expressed in terms of $V_c$, as it corresponds to $2.8 \times 10^{-2} M_{\rm vir}^{0.316}$ \citep{Klypin2011}.
        \item it also preceded \cite{DiCintio2014} in predicting \cite[see][]{DelPopolo2012b} that the inner slope depends on the %ratio of 
        total baryonic content to total mass ratio.%dependence of that inner slope \cite{DelPopolo2012b}
        \item Fig.~6 in \citep{DiCintio2014} compares well with the inner slope change with mass of \citep[Fig. 2]{DelPopolo2016b}.%it compares well with simulations \citep{DiCintio2014} on the inner slope change with mass \citep[Fig. 1]{DelPopolo2016a,DelPopolo2016b}.
        \item it moreover provides a comparison of the Tully-Fisher and Faber-Jackson, $M_{Star}-M_{halo}$, relationships with simulations \citep[Figs. 4, 5]{DelPopolo2016b}.%\citep[Figs. 4, 5]{DelPopolo2016a,DelPopolo2016b}.
\end{enumerate}

\section{Summary of the steps of the simulation}\label{sec:Simulation}
Our model follows a semi-analytic approach, which is inexpensive {compared }with %respect to 
N-body/hydrodynamical simulations ({such as }%like 
NIHAO). This simplifies the construction of samples of galaxies, and rapid exploration of parameter space. Comparison studies of semi-analytic and N-body/hydro simulations have shown a good agreement in the {studied cases }%case of studies case  
\citep[see][and references therein]{Benson2012}.
We use cosmological parameters as given by \cite[Section 2]{Maccio:2020svl}{. Initially, }%, Section 2, and initially 
the system is in gas form with {the ''universal baryon fraction`` \citep{Komatsu2009} }$f_{\rm b}=0.17\pm 0.01$ %for the ''universal baryon fraction`` \citep{Komatsu2009} 
\citep[set to 0.167 in][]{Komatsu2011}. 
The %description of the 
way %the 
initial conditions, starting from the power spectrum, are fixed{, and their ensuing }% and the following 
evolution, is described in \citep[Appendix B]{DelPopolo2009}. When the system reaches the non-linear regime{, }%is calculated the 
tidal interaction with neighbors{ are calculated }%, 
as shown in detail in \citep[Appendix C]{DelPopolo2009}. In the collapse phase, random angular momentum is generated and is calculated as in \citep[Appendix C]{DelPopolo2009}. The effect of dynamical friction is calculated in \citep[Appendix D]{DelPopolo2009}, while %in  of 
\citep[Appendix E]{DelPopolo2009} {shows }%is shown 
how the baryonic dissipative collapse happens. In the collapse the system can give {either }rise to a spiral structure or {to }a spheroid. This is described in \citep[Section A.5]{DelPopolo2016a}. Clumps characteristics and formation {are }%is 
described in Section II. 2 of this paper. Stars form according to the scheme described in Section II. 3 of this paper. The black hole formation, and the AGN feedback {are }%is 
described in the final part of Section II. 3 of this paper.   
{The }%As already reported, the 
NIHAO simulation is a hydrodynamical simulation, based on GASOLINE2: as reported in \citep{Maccio:2020svl} it includes a series of physical effects like
compton cooling, photoionisation and heating from the ultraviolet background, metal cooling, chemical enrichment, star formation and feedback from supernovae
and massive stars.

\section{DFBC and SNF}
\label{sec:DFBC and SNF}

This paper presents DFBC and SNF results within the model of Sec.~\ref{sec:Implementation}. In particular, it focusses on DM halo inner slopes. It will not quantitatively compare those with results from \cite{Maccio:2020svl}, as \begin {enumerate*} [label=\itshape\arabic*.]%\upshape)
                                                                                                                                                                                                                                                \item such comparison was already presented in \citep{DelPopolo2016a} for galaxies, using a large sample of data, against the \citep{DiCintio2014} model, that displays results similar to \cite{Maccio:2020svl}. 
                                                                                                                                                                                                                                                \item the \cite{Maccio:2020svl} model only extends to the mass range of groups, while clusters of galaxies inner slope estimates are available \citep[see][]{Newman2013a,Newman2013b}.
                                                                                                                                                                                                                                               \end {enumerate*}
% % In the present paper we will find the results given by the DFBC and SNF for the inner slope of dark matter halo. We will not compare quantitatively the results of the two models obtained with the model of this paper, and \cite{Maccio:2020svl}. This because 1. this was already done in \citep{DelPopolo2016a} in the case of galaxies, using a large sample of data, comparing with \citep{DiCintio2014} model giving similar results to \cite{Maccio:2020svl}. 2. \cite{Maccio:2020svl} reaches the mass range of groups, and not clusters of galaxies for which there are estimates of the inner slope (see \citep{Newman2013a,Newman2013b}). 
% 
% Such comparison, with clusters mass extrapolation of \cite{Maccio:2020svl} result, will be left for a future paper, considering Milky Way dwarf spheroidals \citep{Hayashi:2020jze} and those \citep{Newman2013a,Newman2013b} clusters, although limitations in density profile inner slope determination are well known. 
% % However, in a future paper we think to compare the two results at with the Milky Way dwarf spheroidals data obtained by \citep{Hayashi:2020jze}, and the \citep{Newman2013a,Newman2013b} clusters, extrapolating 
% % \cite{Maccio:2020svl} result to clusters mass, even if it the limitations in the determination of the inner slope of the density profile are well known. 

The role and importance of baryons in %solving 
the Cusp/Core problem solution was suggested by Flores and Primack \citep{Flores1994}, and by several subsequent %following 
papers. \citep{Navarro1996a} showed that the expulsion of gas in the halo in a single event could flatten the cusp. However, it was soon clear that a single event was not sufficient to produce observed flattening%enough
, and that repeated events were needed \citep{Gnedin2002}. \citep{Mashchenko2006,Mashchenko2008}, showed that random bulk motions of gas due to SN explosion could form a core. Governato \citep{Governato2010,Governato2012} confirmed the result, in addition to finding %and moreover found 
a correlation between stellar mass, $M_\star$, and the inner slope for galaxies with %having
$M_\star > 10^6 M_{\odot}$. Their %The quoted 
simulations, similarly to %that of 
\citep{DiCintio2014}, implement SN feedback through early stellar feedback or the %blast wave 
SN feedback. %As already reported, \citep{DiCintio2014} showed that the inner slope characteristics depends from the ratio stellar mass, halo mass, $M_*/M_{\rm halo}$. This dependence of the slope characteristics from mass 
In fact, the inner slope characteristics dependence on stellar mass to halo mass ratio, $M_\star/M_{\rm halo}$, had already been found by \citep{DelPopolo2010}. 

An alternative mechanism to flatten 
%The other mechanism that can transform 
cusps into cores was proposed by El-Zant \citep{ElZant2001,ElZant2004}. The model is based on %the 
"heating" of DM via %dark matter due to the 
interaction of baryons (gas clumps) with it %dark matter 
through dynamical friction. The exchange of energy and angular momentum between clumps and dark matter can flatten the profile. The earlier the process occurs %happens 
(i.e. for smaller %when 
halos% were smaller
), the more efficient it is. Many studies have confirmed the effectiveness of the process \citep{Ma2004,Nipoti2004,RomanoDiaz2008,DelPopolo2009,RomanoDiaz2009,Cole2011,DelPopolo2014,Nipoti2015}.
In more %More in 
detail, as shown by \citep{DelPopolo2009,DelPopolo2014}, the DM and gas proto-structure starts %containing dark matter and gas, is initially 
in the linear phase. It %The protostructure 
expands to reach a maximum radius, and then recollapse. The collapse of %happens before in 
the DM component occurs first, forming the potential wells in which baryons will fall. Because of radiative processes, baryons form clump, which collapse to the halo centre and form stars. The %One phase of the 
collapse also comprises %is 
the so called "adiabatic contraction" phase \citep{Blumenthal1986,Gnedin2004}, in which baryons are compressed, generating a cuspier DM profiles. 

\begin{figure*}
\begin{centering}
\includegraphics[width=1.4\columnwidth]{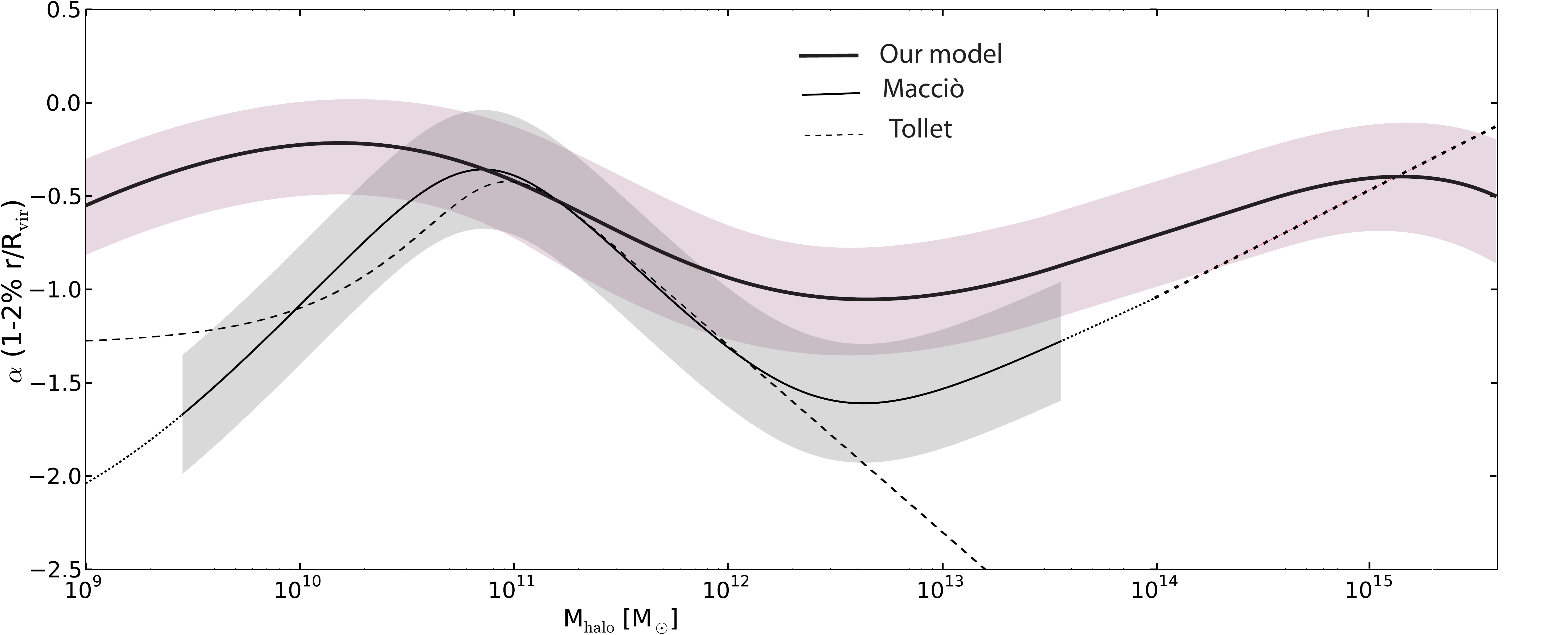}
\par\end{centering}
\caption{\label{fig:NSWD_branches_y01} The inner slope-mass relation for the halo. The top thick line represents the result of this paper, the shaded region the 1-$\sigma$ scatter. The bottom thin line represents the results of \citep{Maccio:2020svl}, and the shaded region the 1-$\sigma$ scatter. The dashed line is the results of \citep{Tollet:2015gqa}. }
\end{figure*}

\begin{figure*}
\begin{centering}
\includegraphics[width=1.4\columnwidth]{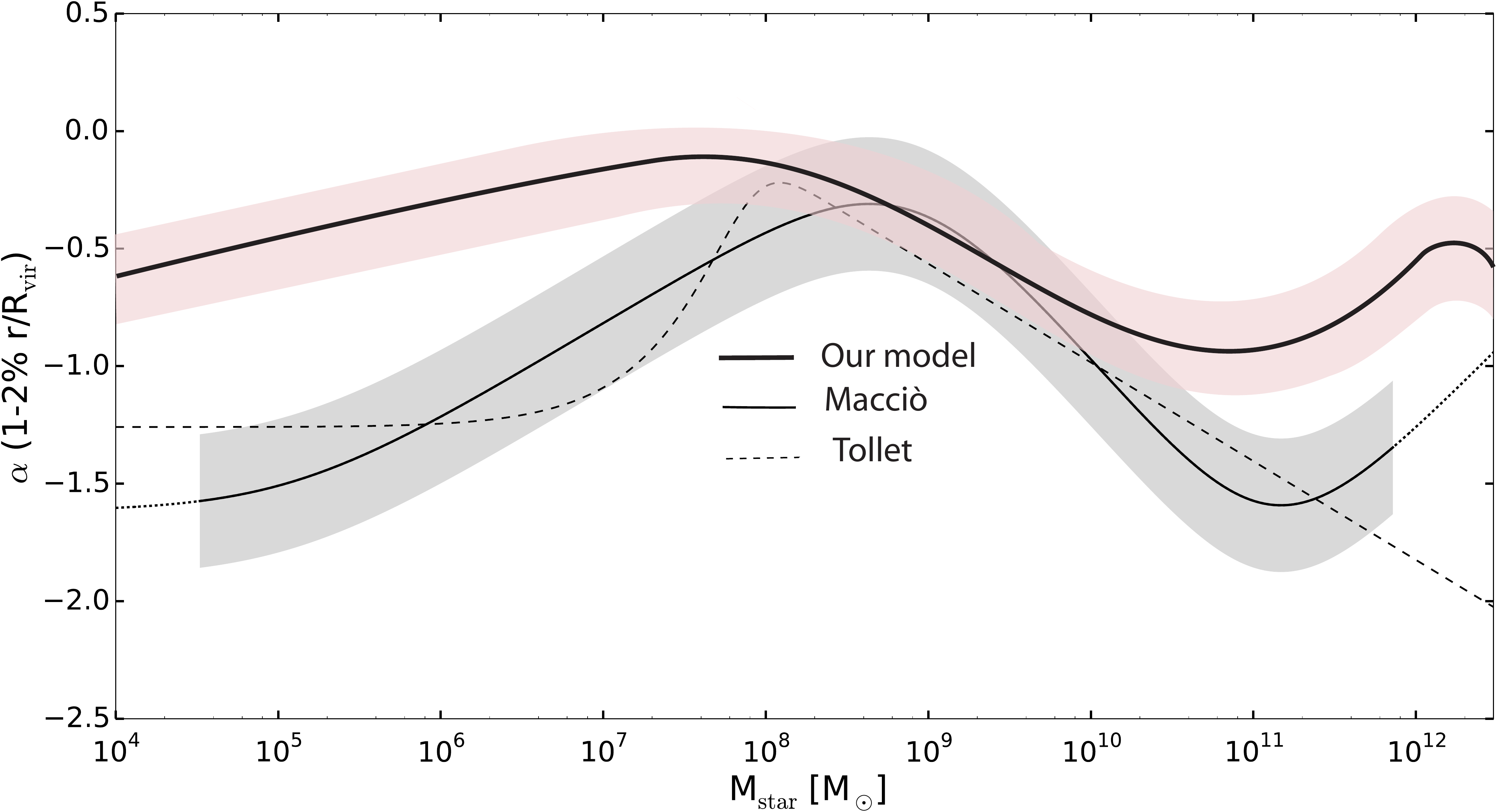}
\par\end{centering}
\caption{\label{fig:einasto} Inner slope-stellar mass relation. Symbols represent the same %clusters and 
models %, %same 
as in Fig.~\ref{fig:NSWD_branches_y01}.}
\end{figure*}

Because of dynamical friction between DM and baryons, clumps fall to the centre of the structure.  During this fall, %In doing this 
angular momentum and energy is transferred to the DM component, and the cusp is heated, giving rise to the formation of a core. This core formation occurs %happens 
before stars formation. Then%After
, stellar feedback expels a large part of the gas, leaving a lower stellar density. After a part of the clumps is transformed into stars, feedback destroys clumps, and mass distribution is dominated by DM. The model just %As told, the model 
described agrees with 
\citep{ElZant2001,Ma2004,Nipoti2004,RomanoDiaz2008,DelPopolo2009,RomanoDiaz2009,Cole2011,DelPopolo2014,Nipoti2015}.

This %is the only 
model is the only one able to describe the correct dependence of the inner slope of the DM density profile from dwarf galaxies to clusters of galaxies \citep{DelPopolo2009,DelPopolo2010,DelPopolo2012a,DelPopolo2012b,DelPopolo2014a}.
Apart from this, as described in the Sec.~\ref{sec:Implementation}, the model  predicted %described 
several results, %that were 
later also obtained %found 
by the SNF model.

The SNF and DFBC model differ significantly in the series of steps they require. The 
SNF model 
starts from gas that forms stars. These can explode into supernovae if they have enough mass. A longer and more complex series of events are needed to produce the observed density profile flattening than for the DFBC. Indeed, the DFBC only requires 
the presence of gas clumps to flatten the halo cusp and gives rise to the core. DFBC %Its flattening 
is therefore 
more ergonomic %\Mov{{\bf[how so?]}and economic }
and efficient at producing cores than the SNF model.

\section{Results}
\label{sec:Results}
We used the model in Sect.~\ref{sec:Implementation} to determine the structure of the objects formed by DM and baryons. The density profile of every object was fitted to obtain the dependence of the inner slope on stellar and DM mass. The fitting method used is similar to that of \citep{Maccio:2020svl}. After determining its %the halo 
center, the halo %it 
is divided in fifty spherical shells, each one with constant width in logarithmic scale. The halo density profile is obtained by evaluating, for each shell, the average DM density. The density profile central slope was then obtained, considering the shells with %We then considered the shells having 
radius in the range 1\%-2\% of the virial radius\footnote{The virial radius is defined as the radius at which the halo overdensity is 200 $\rho_c$, being $\rho_c$ the critical density }. The density profile central slope $\alpha$ was then computed %was fitted 
with a linear fit in the $\log{r}-\log{\rho}$ plane%in order to obtain the profile slope, $\alpha$
.

The main results of the paper are plotted in Figs.~\ref{fig:NSWD_branches_y01}-\ref{fig:einasto}, showing the relations between the inner density slope $\alpha$, and the halo mass, or %and 
stellar mass, respectively. In both %the 
plots, the top solid lines represent the result of this paper, while the bottom ones that of \citep{Maccio:2020svl}. Note that the dotted line in {the \citep{Maccio:2020svl} result is an %model is the relation proposed in NIHAO-IV, based on a subset of the NIHAO galaxies without BH\Anv{, and are 
extrapolation to larger and smaller masses}.

The dashed lines represent the Tollet \cite{Tollet:2015gqa} result. In %Concerning Fig.~\ref{fig:NSWD_branches_y01}, in our model, and in 
the halo mass range, $10^9-2\times 10^{10} M_{\odot}$ of Fig.~\ref{fig:NSWD_branches_y01}, the slope of our model flattens from -0.5 to values closer to zero, %then, after the maximum, it steepens again reaching values of around -1 for masses of $10^{12} M_{\odot}$. 
thus %namely 
there is a maximum of core formation. In the SNF model {\citep[i.e.][the Macci\'o model]{Maccio:2020svl}}%(namely Macci\'o model)
, core formation proceeds from significant alteration of the inner DM density profile from stellar feedback: the inner halo then  expands, giving rise to a core. In the DFBC model (namely our model), the  DF interaction between DM and baryons produces a "heating" of DM, with a consequent expansion, and the formation of a core.

In this 
mass range, Fig.~\ref{fig:NSWD_branches_y01} displays  
a similar trend to that of \citep{Tollet:2015gqa,Maccio:2020svl}%\cite{Tollet:2015gqa}
. 

Although the trends are similar, the slopes differ, %Even if the trend is similar, the slope, 
especially below $10^{11} M_{\odot}$%, differs
. In this range as well as %and also 
in the other ranges, the difference in slope is due to the different ways the DFBC and SNF works. As discussed in the paper, for the DFBC, the flattening of density profiles is due to the ”heating” of DM via interaction with baryons (gas clumps) through dynamical friction. Through this %In the 
interaction, angular momentum and energy are %is 
exchanged between dark matter and clumps, resulting in % with the results of 
the profile flattening. For the SNF, mass ejection from the supernovae leads to the same effect. 
A discriminating issue lies in the onset of the ''heating`` processes: %An important point to recall is that 
the earlier those processes arise%the described processes 
%happen
, the more efficiently they flatten the profile%they are
. 
The series of steps SNF and DFBC require to flatten a profile are different. For the SNF, gas must form stars before %, and 
these can explode as supernovae, if the mass is large enough.
A longer and more complex series of events are needed to produce the observed density profile flattening than for the DFBC. To flatten the halo cusp and give rise to the core, %in the case of 
the DFBC %, is 
just requires %required 
the presence of gas clumps. Its flattening is therefore more ergonomic and efficient than the SNF model. As a result the process produces smaller slopes than through SNF.

After the maximum, both our result and that of Macci\'o steepen again, reaching values %of 
around -1, in our model, and -1.6 for Macci\'o model, for masses of $4 \times 10^{12} M_{\odot}$.

That steepening, especially for the Macci\'o model based on SNF, is related to stars forming in the central regions, which deepens the gravitational potential, opposing SN feedback, and the DM expansion process. This produce a cuspier profile.

In the mass range above %larger than 
$4\times10^{12}%-\times 10^{13} 
M_{\odot}$, the slope starts to flatten.%continues to steepen,  

In the case of the Macci\'o model, the flattening stops at the limit of their simulation, namely at $3.6 \times 10^{13}$. In our case, the flattening reaches a maximum at $\simeq 10^{15} M_{\odot}$, and %the flattening 
is related to the AGN feedback. This is similar to the flattening effect of SN feedback on smaller masses. Beyond the %that 
$\simeq 10^{15} M_{\odot}$ maximum, we observe again a steepening of the profile, because AGN feedback  becomes less effective.

%then invert its behavior, flattening until masses of $10^{15} M_{\odot}$ are reached. Namely, a relaxation of the halo is observed, with an inner slope flatter than the DM only N-body simulations predictions. The 
%slope steepens again for masses larger than that $10^{15} M_{\odot}$ limit. 
In Summary, while in \citep{DelPopolo2016a}, and \cite{Tollet:2015gqa}, the behavior is non-monotonic with only one %just }a 
maximum of "core-formation"% in a mass range around  
%$2 \times \simeq 10^{10} M_{\odot}$
, %now in our model 
the behavior in our model is more complex, and presents %are present 
two maxima of "core-formation". The first maximum is %one 
produced by the DFBC mechanism and SN feedback, while %and 
the second is %one 
produced by the DFBC mechanism and AGN feedback. The situation in the case of \citep{Maccio:2020svl} is similar to that of \cite{Tollet:2015gqa} in the halo mass range $\simeq 2 \times 10^{9} M_{\odot}-10^{12} M_{\odot}$, where %and %. Namely, 
their slope flattens to a maximum  
at $\simeq 10^{11} M_{\odot}$. 
A relaxation of the halo is observed for 
masses larger than $\simeq 2 \times 10^{12} M_{\odot}$, as in the case of our model, 
with an inner slope flatter than the predictions of DM only N-body 
simulations. 

At this point there are two important remarks. The first concerns an effect in the Macci\'o model that prevents it from producing shallow enough cores in the mass range $\simeq 10^{12}-10^{13} M_{\odot}$: %Here, we want to stress two points. First, in the case of the Macci\'o model, 
%as we already reported before, in this mass range ($\simeq 10^{12}-10^{13} M_{\odot}$) 
the number of stars forming in the central regions is so large that it can efficiently oppose the SN feedback. This produces a region with steeper slopes, with respect \Mov{to }the region $\simeq 10^{11}-10^{12} M_{\odot}$. This effect is not so important in the case of our model, and consequently we have shallower %smaller 
slopes. The second provides an extra reason for %Second, another reason giving rise to 
the slope difference in this region between our model and Macci\'o's: the larger %increased 
%is due to the already described larger 
efficiency of the DFBC model compared with the SNF model, that does not need to wait for stars to form to start producing %(as the SNF model) to have a 
feedback on DM.

Our model and \citep{Maccio:2020svl} have similar behavior until
$\simeq 3.6 \times  10^{13} M_{\odot}$. As the Maccio model \citep{Maccio:2020svl} validity does not extend beyond $\simeq 3.6 \times 10^{13} M_{\odot}$, 
the comparison with     
our model cannot be extended to some of the masses it reaches.

\begin{table}
\begin{tabular}{lcccccc}
\hline 
{\scriptsize{}$x$ {[}M$_{\odot}${]} }  & {\scriptsize{}$n$ }  & {\scriptsize{}$n_{1}$}  & {\scriptsize{}$x_{0}$ {[}M$_{\odot}${]} }  & {\scriptsize{}$x_{1}$ {[}M$_{\odot}${]} }  & {\scriptsize{}$x_{2}$ {[}M$_{\odot}${]} }  & {\scriptsize{}$x_{3}$ {[}M$_{\odot}${]} } \tabularnewline
{\scriptsize{}$M_{{\rm halo}}$ }  & {\scriptsize{}-5.32 }  & {\scriptsize{}8.60 }  & {\scriptsize{}$1.50\cdot10^{25}$ }  & {\scriptsize{}$2.53\cdot10^{12}$ }  & {\scriptsize{}$3.52\cdot10^{-5}$ }  & {\scriptsize{}$2.49\cdot10^{10}$ } \tabularnewline
\hline 
{\scriptsize{}$M_{\star}$ }  & {\scriptsize{}-0.46 }  & {\scriptsize{}3.94}  & {\scriptsize{}$8.60\cdot10^{12}$ }  & {\scriptsize{}$9.34\cdot10^{10}$ }  & {\scriptsize{}$8.69\cdot10^{2}$ }  & {\scriptsize{}$1.27\cdot10^{8}$ } \tabularnewline
\hline 
 & {\scriptsize{}$\beta$ }  & {\scriptsize{}$\gamma$ }  & {\scriptsize{}$\delta$ }  & {\scriptsize{}$\sigma$} &  & {\scriptsize{}1-$\sigma$}\tabularnewline
{\scriptsize{}$M_{{\rm halo}}$ }  & {\scriptsize{}1.14 }  & {\scriptsize{}0.26}  & {\scriptsize{}0.43 }  & {\scriptsize{}1.15} &  & {\scriptsize{}0.32}\tabularnewline
\hline 
{\scriptsize{}$M_{\star}$ }  & {\scriptsize{}1.16}  & {\scriptsize{}2.53}  & {\scriptsize{}0.22 }  & {\scriptsize{}0.71} &  & {\scriptsize{}0.28}\tabularnewline
\hline 
\end{tabular}{\scriptsize{}\caption{\label{table}Parameters values for the fitting functions described
in eqs.~\ref{eq1}.}
} 
\end{table}
The 
behavior of the inner slope $\alpha$ versus the stellar mass (Fig. (\ref{fig:einasto})), $M_\star$ %, the situation 
is similar to that discussed for the case of $M_{\rm halo}$: %. So, 
in our model, we observe the double maxima, as in the case of $M_{\rm halo}$, at $M_\star \simeq 4 \times 10^{7} M_{\odot}$, and 
$M_\star \simeq 10^{12} M_{\odot}$. For %In the case of 
the \citep{Maccio:2020svl} model, the situation of $M_\star$ is similar to that of $M_{\rm halo}$. 
%This is to be contrasted with the model from \cite{Tollet:2015gqa}, where 
In the \citep{Maccio:2020svl} model, an up-turn for $M_\star > 10^{11} M_{\odot}$ is observed. The behavior of the inner slope as function of $M_{\rm halo}$, or 
$M_\star$, can be represented by the functional forms

\begin{align}
%\hspace{-1.6cm}
\alpha_{\star/{\rm halo}}\left(x\right)  = & n-\log_{10}\left[n_{1}\left(1+\frac{x}{x_{1}}\right)^{-\beta}+\left(\frac{x}{x_{0}}\right)^{\gamma}\right]\nonumber \\
 &   +\log_{10}\left[1+\left(\frac{x}{x_{2}}\right)^{\delta}\right]-\log_{10}\left[1+\left(\frac{x}{x_{3}}\right)^{\sigma}\right]\ . \label{eq1}
\end{align}

%The values 
Note the functional forms are identical but differ from the values of the functional break limit masses ($10^{14}\mbox{M}_{\odot}$ or $10^{12}\mbox{M}_{\odot}$), of the variables ($M_{\mathrm{halo}}$ or $M_{{\star}}$) and 
of the parameters, that are shown in Table \ref{table}.

% \begin{table*}
% \begin{tabular}{lccccccccc}
% \hline
% $x$  [M$_{\odot}$] & $n$ & $n_1$ & $x_0$ [M$_{\odot}$] &  $x_1$ [M$_{\odot}$] &  $x_2$ [M$_{\odot}$] & $\beta$ & $\gamma$ & $\delta$ & 1-$\sigma$\\
%  \hline
% $M_{\rm halo}$ & -0.38 & 9.63 & $1.10\cdot 10^{11}$  & $1.01\cdot 10^{7}$  & $2.26\cdot 10^{12}$ & 0.42 & 0.72 & 1.09 & 0.32?\\
% $M_{\rm star}$ & -0.87 & 0.34 & $1.58\cdot 10^{10}$  & $1.12\cdot 10^{6}$  & $1.16\cdot 10^{11}$ & 0.39 & 0.41 & 1.09 & 0.28?\\
% %
% %\hline
% %\hline
% %$x$ & $n$ & $n_1$ & $n_2$ & $x_0$ &  $x_1$  & $\beta$ & $\gamma$ &  $\epsilon$ &  1-$\sigma$\\
% % \hline
% % $M_\mathrm{stars}/M_\mathrm{halo}$ & -0.0385 & 39.11 & -2.58 & $7.51\cdot 10^{-3}$  & 
% %$5.12\cdot 10^{-5}$  & 0.728 & 1.84 &  0.708 & 0.37 and 0.28\\
% \hline
% \end{tabular}
% \caption{\label{table2}Parameters values for the fitting functions described in eq. \ref{eq1} and \ref{eq2}.}
% \end{table*}

%In 
Figs.~\ref{fig:NSWD_branches_y01}-\ref{fig:einasto} %we draw 
also presents the scatter around both our and the \citep{Maccio:2020svl} %all 
relations. These scatters %They 
are almost constant, and in the case of the slope-halo mass relation, the value of the average scatter is $\sigma \simeq 0.3$, while in the case of the slope-stellar mass relation it is $\sigma \simeq 0.27$. 
It %The scatter 
was calculated using all the galaxies %that we 
simulated. We do %preferred 
not %to 
plot all the galaxies so as to keep the figures legible%to have a plot not too full of symbols
.

Note that the slope behaviour, and by extension that of the DM density profile, for cluster-type masses can result from a model with two stages. The first dissipative phase sees the formation of the seed for the BCG, while the second, dissipationless stage is driven by the DF, between DM in the halo and the sunk baryonic clumps (to the centre), into flattening the density profile inner slope. 

The large scatter of the inner slope among the cluster population reflects \begin{enumerate*}[label=\itshape\arabic*.\upshape)]
    \item that the total mass density profile is given by the sum of the DM and baryon contents: $M_{Total}=M_{BCG}+M_{DM}$ 
    \item haloes follow NFW-like density profiles
    \item the variations in the BCGs masses.
\end{enumerate*}
This entails DM profile inner slopes ranges from the NFW slope to flatter slopes, depending on the amount of central baryons.

Before concluding, let us emphasize %we want to give a final description of 
our results, in their %and the 
differences with \cite{Tollet:2015gqa,Maccio:2020svl}. The major difference lies in the %To start with, our model describes the slope in a 
larger range, from dwarf galaxies to clusters, described by our model, compared with %than 
\cite{Tollet:2015gqa,Maccio:2020svl}%, from dwarf galaxies to clusters
. Indeed, %Looking at 
Fig.~(\ref{fig:NSWD_branches_y01}) reveals that our model produces a flatter slope %, we see that the slope of our model is flatter 
than %that of 
\cite{Tollet:2015gqa,Maccio:2020svl} in all mass ranges, except for %in 
a very small mass range close to $10^{11} M_{\odot}$. {Our model's almost always shallower slope compared with }%The slope predicted by our model is flatter than 
\cite{Tollet:2015gqa,Maccio:2020svl} stems from the more efficient flattening from the DFBC model, compared with the SNF model%because our model is based on the DFBC model which is more efficient than the SNF
. This is particularly visible in the mass range $10^{9}-10^{10} M_{\odot}$. 
Close to $10^{11} M_{\odot}$, %the SNF has its maximum efficiency, because 
the energy released by supernovae is larger than the gravitational potential due to stars,  resulting into a maximum efficiency of the SNF and a corresponding minimum (maximum flattening) for the %and the slopes of 
\cite{Tollet:2015gqa,Maccio:2020svl} %has a minimum (maximum flattening), and the 
slope, while our model outputs similar slopes% in our model is close to that in \cite{Tollet:2015gqa,Maccio:2020svl}
. The intensity of the stars gravitational potential increases from $10^{11} M_{\odot}$ to $5 \times 10^{12} M_{\odot}$, consequently steepening the slope %and as a consequence 
in the \cite{Tollet:2015gqa,Maccio:2020svl} model% the slope steepens
. This behaviour occurs similarly in %is what 
%happens \Anv{in }%also to 
our model, due to the decrease in %because 
the exchange of energy between clumps and DM. It produces a shallower slope, reflecting the higher % decreases. 
%However, as already written, the slope in our model is flatter, because of the larger 
efficiency of the DFBC mechanism compared with %on 
the SNF, as pointed above. In the %. The 
mass range $5 \times 10^{12}- 3.6 \times 10^{13} M_{\odot}$,
AGN feedback %starts to have its 
%effect 
starts to show its effect, %producing the flattening 
in the \cite{Maccio:2020svl} model and in ours, resulting in the observed flattening.

\section{Conclusions% and discussions
}
\label{sec:Conclusions}
Despite the facts that the $\Lambda$CDM model has shown many observational successes and that the Cusp/Core problem is better understood compared with a couple of decades ago, the %The 
Cusp/Core problem remains one of the prominent problems of the $\Lambda$CDM model%, despite the latter model otherwise many observational successes and the former problem's better understanding compared with a couple of decades ago
. It consists in the discrepancy between the inner slope observed in dwarf galaxies, and the cuspy profiles obtained in N-body only simulations. One of its remaining issues concerns the understanding of the observed variations of inner slopes among different kinds of galaxies, as well as among galaxy clusters, with shallower than the standard Navarro-Frenk-White profiles. Following the mass dependence of galaxies inner slopes shown in \citep{DelPopolo2010}, several SPH simulations studied the problem in detail.

%  }
% \Mov{The $\Lambda$CDM model is succesful in many observations, but it shows some issues at scale of galaxies. One of this problem is the Cusp/Core problem, namely the discrepancy between the inner slope observed in dwarf galaxies, and the cuspy profiles obtained in 
% N-body only simulations. Nevertheless this problem is better understood than a couple of decades ago, some points still remain open. One of this points is to understand why different kind of galaxies show different inner slopes, and why some clusters have profiles flatter that the Navarro-Frenk-White profile. After \citep{DelPopolo2010} showed that the inner slope of galaxies depends on mass several SPH simulations have studied the problem in detail.

Following the 
slope-mass relation 
obtained by \citep{DiCintio:2014xia,Tollet:2015gqa,DelPopolo2016a} in the mass range covering 
dwarf galaxies up to Milky-Way sizes, and its extension  
by \citep{Maccio:2020svl} to galaxy groups, 
this paper  
further extended the 
\citep{Maccio:2020svl} results to the mass range of %to 
clusters of galaxies% masses
. For more massive structures %masses %larger 
than %those of 
galaxies, the inner slope-mass relation continues to steepen until %till 
reaching a minimum %, 
and increasing again. Specifically, %then invert its behavior. In the case of 
the slope-halo mass profile %, the 
flattening starts at masses $\simeq 10^{12.4} M_{\odot}$
and reaches a maximum of core formation at $\simeq 10^{15} M_{\odot}$. This %The 
flattening is produced by the action of AGN feedback, in a similar way to the role of %similarly to what happened at smaller masses for the 
SN feedback for smaller galaxy masses. Beyond this %.
%After reaching the quoted 
maximum of core formation, the trend reverts to %changes again and we observe a 
steepening. The one-sigma scatter on $\alpha$ is approximately constant in %all
the whole mass range ($\Delta\alpha\simeq 0.3$). 
Our %The 
slope-mass relation is a first step in determining a density profile taking into account baryons, for a larger mass range than for the profiles %as that 
obtained by \citep{DiCintio:2014xia}%, but for a larger mass range
. 
Indeed, the \citep{DiCintio:2014xia} %obtained a 
density profile taking %account of 
baryons into account is limited to %onlt in 
the %mass range 
dwarf galaxies to %, 
Milky Way size mass range%like galaxies
. We have extended %will extend 
this to clusters of galaxies.
In a subsequent paper, we propose to find the density profile of structures, taking into account the role of baryons, in the mass range from dwarf galaxies to clusters. Despite well known limitations in density profile inner slope determination, such model could then be compared with clusters mass extrapolation of the \cite{Maccio:2020svl} results, also considering Milky Way dwarf spheroidals \citep{Hayashi:2020jze} and clusters from \citep{Newman2013a,Newman2013b}.

% Such comparison, with clusters mass extrapolation of \cite{Maccio:2020svl} result, will be left for a future paper, considering Milky Way dwarf spheroidals \citep{Hayashi:2020jze} and those \citep{Newman2013a,Newman2013b} clusters, although limitations in density profile inner slope determination are well known. 

\section*{Acknowledgments}
%This is the most common positions for acknowledgments. A macro is available to maintain the same layout and spelling of the heading.
%\paragraph{Note added.} This is also a good position for notes added after the paper has been written.

MLeD acknowledges the financial support by the Lanzhou University starting
fund, the Fundamental Research Funds for the Central Universities
(Grant No. lzujbky-2019-25), National Science Foundation of China (grant No. 12047501) and the 111 Project under Grant No. B20063. %M.D. thanks Xin Wu for his support
%in this research during the COVID-19 quarantine measures.
The authors wish to thank Maksym Deliyergiyev for some calculations.
%\end{acknowledgements}

%%%%%%%%%%%%%%%%%%%%%%%%%%%%%
%%%%%%%%%%%%%%%%%%%%%%%%%%%%%
% \appendix
% g
\bibliographystyle{apsrev4-1}
%\bibliography{unified.bbl}
\bibliography{old_MasterBib2,biblioNS}
\end{document}